\documentclass{achemso}

\usepackage[utf8]{inputenc}
\usepackage{physics}
\usepackage{graphicx}
\usepackage{here}
\usepackage{bm}
\usepackage{color}
\usepackage{parskip}

\title{Theoretical analysis of chemical reactions using a variational quantum eigensolver method without specifying molecular charge}

\author{Soichi Shirai}
\email{shirai@mosk.tytlabs.co.jp}
\author{Takahiro Horiba}
\author{Hirotoshi Hirai}
\affiliation{Toyota Central Research and Development Laboratories, Incorporated,\\41-1 Yokomichi, Nagakute, Aichi 480-1192, Japan}

\begin{document}

\newpage

\section{Abstract}
Quantum chemical calculations have attracted much attention as a practical application of quantum computing.
Quantum computers can prepare superpositions of electronic states with various numbers of electrons on qubits.
This special feature could be used to construct an efficient method for analyzing the structural variations of molecules and chemical reactions involving changes in molecular charge.
The present work demonstrates a variational quantum eigensolver (VQE) algorithm based on a cost function (\textit{L}$_{\text{cost}}$) having the same form as the grand potential of the grand canonical ensemble of electrons.
The chemical potential of the electrons (\textit{w}) is used as an input to these VQE calculations, whereas the molecular charge is not specified in advance but rather is a physical quantity that results from the calculations.
Calculations involving model systems are carried out to show the viability of this new approach.
Calculations for typical electron-donating and electron-accepting molecules using this technique yielded cationic, neutral or anionic species depending on the value of \textit{w}.
Models representing the adsorption of water or ammonia on copper-based catalysts predicted that oxidation would be associated with such adsorption. The molecular structures in which such reactions occurred were found to be dependent on the catalyst model, the adsorbed molecular species, and the value of \textit{w}.
These results arise because the electronic state that gives the lowest \textit{L}$_{\text{cost}}$ value depends on the value of \textit{w} and the molecular structure.
This behaviour was successfully simulated by the present VQE calculations.

\newpage

\section{Introduction}

Quantum chemical calculations are widely used in materials research to analyze and predict molecular structures and chemical reactions by calculating the wave functions that determine the behavior of electrons.~\cite{gagliardi2015, ionic-liquid-2017, mennucci2017, dft2022, maeda2023}
Recently, quantum chemical calculations have attracted much attention as a practical application of quantum computing.~\cite{ijqcrev, qcqc2019, qcqc2020-1, qcqc2020-2}
In the configuration interaction (CI) method, a typical post-Hartree-Fock technique, the wave function is constructed as a linear combination of the Slater determinants corresponding to the electron configurations.~\cite{ci1999}
The CI method using all possible configurations is referred to as the full CI approach and provides an exact solution for the basis set under consideration.~\cite{fci1996}
Unfortunately, the number of possible electron configurations increases rapidly (according to a factorial relationship) with respect to the number of electrons and orbitals, such that the full CI method is applicable only to very small systems.
For this reason, the so-called truncate CI technique, which uses only the large contributing configurations, and density functional theory (DFT) calculations with more reasonable computational costs are more typically employed.
Quantum computers based on qubits have the potential to perform full CI calculations in polynomial time by using quantum phase estimation (QPE) algorithms. These devices are expected to allow highly accurate quantum chemical calculations to be performed that cannot be achieved by conventional methods.~\cite{qpe1999, qpe2005, qpe2014, qpe2017} 
At present, quantum computers comprise noisy intermediate-scale quantum (NISQ) devices based on qubits without error corrections and so cannot handle the deep quantum circuits required for quantum chemical calculations.~\cite{nisq2018, nisq2019, nisq2021}
For this reason, a quantum-classical hybrid algorithm referred to as the variational quantum eigensolver (VQE) has been widely adopted to perform quantum chemical calculations. In the case of this algorithm, expected values are evaluated by an NISQ whereas variational  parameters are optimized using a classical computer.~\cite{vqe2014, vqe2016, vqe2022fedorov}

In the case of quantum chemical calculations based on quantum algorithms, superpositions of electronic states with different expected values for various physical quantities can be prepared on qubits.
This unique feature enables calculations to be executed differently compared with traditional methods, which could offer extra advantages in addition to the computational scaling of highly correlated systems, a potential benefit of quantum computing.
As an example, when using the VQE algorithm, which utilizes superpositions of electronic states having different spin multiplicities, the calculation converges to the lowest energy ground state regardless of spin multiplicity. Hence, there is no need to specify the spin multiplicity in advance or to perform calculations for each spin multiplicity and compare the results, as is required in conventional calculations.~\cite{vqeptco}
In contrast, the spin multiplicity is determined as a result of the calculation.
This approach may be applied to the analysis of chemical reactions involving changes in spin multiplicity~\cite{kitagawa2016dft, watanabe2016spin, jiang2016mechanism, kwon2017catalytic, Nakatani2018, li2020identification} and to the calculation of molecular systems in which the ground-state spin multiplicity is non-trivial~\cite{radon2008binding, sebetci2009does, ali2012electronic, garcia2017effect, zhang2020probing}.
Such calculations can be performed in a more efficient manner than is possible when using a conventional framework.

The present study demonstrates VQE calculations that do not specify the molecular charge.
The VQE method provides the state with the lowest given cost function (\textit{L}$_{\text{cost}}$) by optimizing the variational parameters.
When adopting an ansatz that does not preserve the number of electrons, a penalty term is used to specify the number of electrons and to force convergence from a superposition of states with different charges to a state having the specified charge. 
The present approach also utilizes a superposition of electronic states having different charges but assumes that \textit{L}$_{\text{cost}}$ is of the same form as the grand potential,~\cite{aimd-sprik, bonnet2012, jcp-gce} which is the thermodynamic quantity in the grand canonical ensemble for the electron number.
On this basis, it is possible to carry out VQE calculations that allows the electron number to vary in conjunction with a constant chemical potential for the electron.
This method may be applied to the analysis of chemical reaction mechanisms involving charge changes at a specified potential and to the search for electrochemical catalysts that can make such reactions more efficient.
As an example, molecular charges were calculated as functions of potential for typical electron-donor and electron-acceptor molecules.
Calculations were also performed based on modeling the adsorption of water and ammonia on copper-containing catalysts, which is the first step in the oxidation reactions of these molecules.

\section{Theory}
The VQE cost function \textit{L}$_{\text{cost}}$ is defined as~\cite{vqe2014, vqe2016}

\begin{equation}
    \label{lcost1}
    L_\text{cost}=\bra{\Psi_o}
    \hat{U}^{\dagger}(\bm{\theta}) \hat{H}\hat{U}(\bm{\theta})\ket{\Psi_o}+L_\text{penalty},
\end{equation}

where $\ket{\Psi_o}$ is the initial function, $\hat{U}(\bm{\theta})$ is the quantum circuit with \bm${\bm{\theta}}$ as a variational parameter, and $\hat{H}$ is the Hamiltonian.
Here, \textit{L}$_{\text{penalty}}$ is the penalty term that increases \textit{L}$_{\text{cost}}$ in the case that the expected value of the physical quantity for the wave function deviates from the specified value.
For a VQE algorithm with an associated ansatz that does not preserve the number of electrons, convergence to a state for which the molecular charge matches the specified value can be obtained from

\begin{equation}
    \label{lpenalty1}
    L_\text{penalty}=w\bra{\Psi_o}
    \hat{U}^{\dagger}(\bm{\theta}) (\hat{N}-n_{p})^2
    \hat{U}(\bm{\theta})\ket{\Psi_o},
\end{equation}

where $\hat{N}$ is the electron number operator, which returns the electron number \textit{n} as an expectation value when acting on the wave function, \textit{n$_{p}$} is a specific number of electrons used as an input value, and \textit{w} is a weighting coefficient that determines the magnitude of the penalty.
Note that, in the case that the eigenvalue for $\hat{N}$, \textit{n}, matches \textit{n$_{p}$}, the penalty term will be zero.
If \textit{n} does not equal \textit{n$_{p}$}, the penalty term has a nonzero value and \textit{L}$_{\text{cost}}$ is increased.
Therefore, by minimizing \textit{L}$_{\text{cost}}$ based on employing \textit{L}$_{\text{penalty}}$ in eq \ref{lpenalty1}, we can obtain a wave function for which \textit{n} is consistent with \textit{n$_{p}$}.
In this manner, the penalty term forces convergence to the desired state from a superposition of electronic states prepared on the qubits and the solution obtained depends on the penalty term.

In a typical quantum chemical calculation, \textit{n} is a constant because the charge of the molecule is specified (canonical ensemble with respect to the number of electrons).
In contrast, superpositions of electronic states with different numbers of electrons have potential to enable the calculation without specifying a constant number of electrons by combining appropriately prepared penalty terms.
Here, we focus on the grand canonical ensemble for the electron number.
In the case of the grand canonical ensemble approach, the chemical potential of the electrons (\textit{$\mu_{\text{e}}$}) is constant, whereas \textit{n} is variable.
The thermodynamic potential under these conditions can be described by the grand potential ($\Omega_{\text{e }}$), defined as~\cite{aimd-sprik, bonnet2012, jcp-gce}

\begin{equation}
    \label{gp1}
    \Omega_{\text{e}} = A - \mu_{\text{e}} \cdot n,
\end{equation}

where \textit{A} is the Helmholtz free energy. Accordingly, \textit{L}$_{\text{penalty}}$ in eq~\ref{lcost1} can be modified to give

\begin{equation}
    \label{lcost2}
    L_\text{cost} = \bra{\Psi_o}\hat{U}^{\dagger}(\bm{\theta}) \hat{H}\hat{U}(\bm{\theta})\ket{\Psi_o}
    -w\bra{\Psi_o}\hat{U}^{\dagger}(\bm{\theta}) \hat{N}\hat{U}(\bm{\theta})\ket{\Psi_o}.
\end{equation}

Replacing the first and second terms in eq~\ref{lcost2} with $E(\bm{\theta})$ and $w \cdot n(\bm{\theta})$, respectively, gives

\begin{equation}
    \label{lcost3}
    L_\text{cost} = E(\bm{\theta})-w \cdot n(\bm{\theta}).
\end{equation}

This relationship has the same form as eq~\ref{gp1} when assuming a temperature of 0 K and allows us to map \textit{w} to the chemical potential of the electron.
Using eq~\ref{lcost3} to define \textit{L}$_{\text{cost}}$, VQE calculations for variable numbers of electrons can be performed in conjunction with a constant electron chemical potential of \textit{w}.
It should be noted that the molecular charge is determined from these calculations and is not specified a priori.
Although a change in the number of electrons is accompanied by a change in spin multiplicity, eq~\ref{lcost3} does not include a penalty term related to spin multiplicity.
Therefore, the change in spin multiplicity associated with a change in the number of electrons is allowed without an increase in \textit{L}$_{\text{cost}}$.

In a typical VQE calculation, a limited number of active orbitals and active electrons will be selected and so the number of inactive electrons does not change from state to state.
Therefore, the difference in \textit{L}$_{\text{cost}}$ between states will be the same if \textit{n} is the number of active electrons rather than the total number of electrons.
This approach was adopted in the following calculations.

\section{Computational Details}
Molecular geometric structures were optimized using the MP2~\cite{mp2} method with the def2-TZVP basis set~\cite{def2}. The VQE calculations were performed using the def2-SVP basis set.~\cite{def2}
A hardware-efficient ansatz~\cite{HE-ansatz} was used as a parameterized quantum circuit with a circuit depth of two.
The Jordan-Wigner transformation~\cite{jw1, jw2} was used to map the Hamiltonian to the quantum circuit. The adaptive moment estimation (ADAM)~\cite{adam} technique was employed to optimize $\bm{\theta}$.
All VQE calculations were performed using a noise-free quantum circuit simulator.
The neutral, oxidized and reduced states of each molecule were calculated using the complete active space CI (CASCI) method to obtain reference values.~\cite{casci}
The molecular structures, molecular orbitals, basis sets and active orbitals employed during these CASCI calculations were the same as those used in the VQE calculations.
The molecular structure optimizations and CASCI calculations were performed using the GAMESS~\cite{gamess1993, gamess2005, gamess2020} software package whereas the VQE calculations were carried out with a program prepared using Pennylane, a cross-platform Python library provided by Xanadu.~\cite{pennylane}

\subsection{Calculations involving typical electroactive organic molecules}
Pyrrole and thiophene are widely used as building units for the synthesis of hole transport materials,~\cite{polypyrrole1, polythiophene1, polythiophene2} while ~\textit{p}-benzoquinone is a typical electron-accepting molecule ~\cite{quinone1}.
Hence, the present work analyzed variations in the charges of these molecules while changing the value of ~\textit{w} in eq \ref{lcost3}.
After optimizing the molecular structure of each compound, VQE calculations were carried out using eq \ref{lcost3} to determine the cost function while varying \textit{w} over the range from -15.0 to +10.0 eV.
During these calculations, the Hartree-Fock molecular orbitals obtained for the singlet neutral state of each molecule were employed.
The highest occupied molecular orbital (HOMO) and the lowest unoccupied molecular orbital (LUMO) were set as the active orbitals in each case and four qubits were used for the VQE calculations.

\subsection{Models for the adsorption of ammonia and water on Cu and CuO}
The oxidation of water (2H$_{2}$O $\rightarrow$ O$_{2}$ + 4H$^{+}$ + 4e$^{-}$) is an essential component of artificial photosynthesis, which converts solar energy into storable chemical energy in conjunction with the reduction reaction of carbon dioxide.~\cite{karkas2014, ye2019water}
Ammonia, which has a high hydrogen content, is a promising hydrogen carrier.~\cite{zhai2023ammonia}
Despite this, the use of ammonia as a hydrogen carrier requires the oxidation of ammonia to extract hydrogen (2NH$_{3}$ $\rightarrow$ N$_{2}$ + 3H$_{2}$).
A highly durable electrochemical catalyst that can carry out these reactions with high efficiency is required.
Catalysts using the earth-abundant metal copper have received significant attention in this regard.
In particular, copper complexes with specific ligands~\cite{cu-complex-water1, cu-complex-water2, cu-complex-ammonia} and solid materials based on copper oxide~\cite{cuo-water-2017, cuo-water-2021, cuo-ammonia-2022} are known to exhibit catalytic activity.
The adsorption of these molecules on Cu and CuO can serve as a simple model to describe the first step of a reaction catalyzed by copper-containing materials.

\begin{figure}[H]
    \begin{center}
        \includegraphics[scale=1.0]{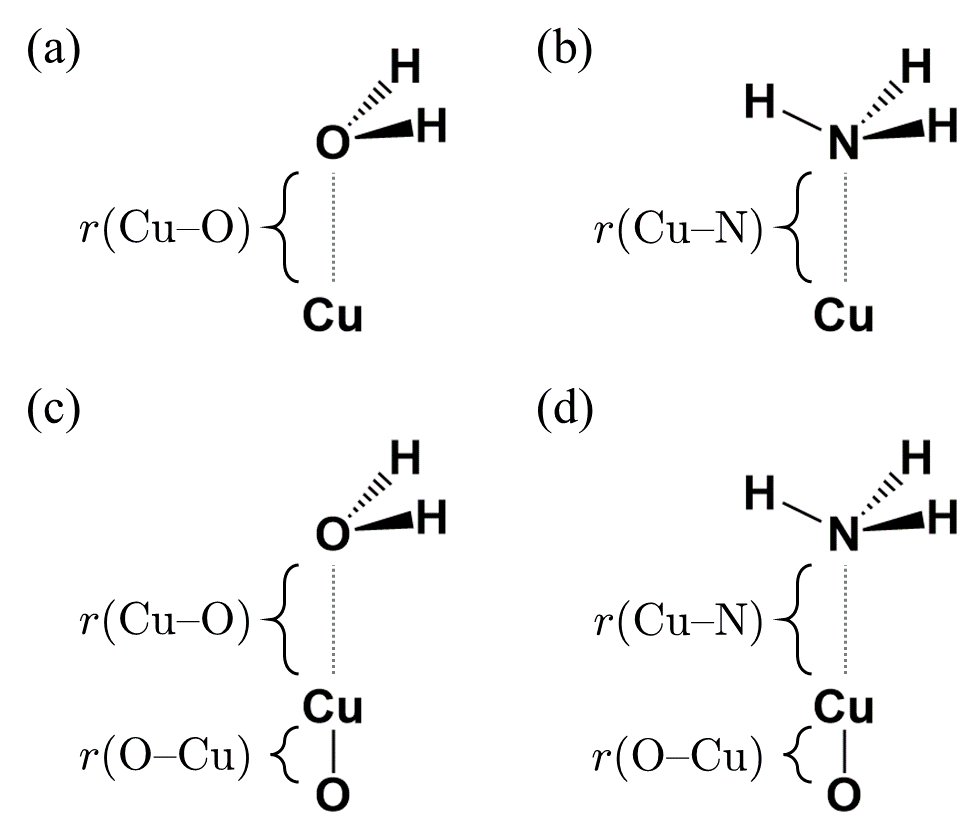}
        \caption{Molecular structures of the models used for calculations: (a) Cu--(H$_{2}$O), (b) Cu--(NH$_{3}$), (c) CuO--(H$_{2}$O), and (d) CuO--(NH$_{3}$).}
        \label{3models}
    \end{center}
\end{figure}

The molecular structures of the models for which the calculations were performed are shown in Figure~\ref{3models}.
The structural optimization calculations for the neutral state with a molecular charge of $\pm$0 resulted in convergence to structures having $C_{s}$ symmetry in the case of the Cu--(H$_{2}$O) and CuO--(H$_{2}$O) and $C_{3v}$ symmetry in the case of the Cu--(NH$_{3}$) and CuO--(NH$_{3}$).
In subsequent calculations, these symmetries were maintained and the bond lengths, bond angles and dihedral angles were fixed with the exception of \textit{r}(Cu--L) (L = O, N).
The potential energy curves for \textit{r}(Cu--L) values of 1.8 -- 5.0 \AA \ were calculated using eq \ref{lcost3} as the VQE cost function while varying the weighting coefficient \textit{w} for Cu--X and CuO--X (X = H$_{2}$O, NH$_{3}$).
This value was varied over the range of $-6.4$ -- $-4.4$ eV for Cu--X and over the range of $-8.0$ -- $-6.9$ eV for CuO--X.
Hartree-Fock orbitals obtained for the +1 oxidation state, meaning a closed-shell singlet, were used as the molecular orbitals. The highest energy Cu 4s and O 2p or N 2p orbitals were used as the active orbitals.
Calculations were also performed for models for which \textit{r}(Cu--L) = 20.0 \AA \ to examine the dissociation limit of each system.

\newpage

\newpage
\section{Results and Discussion}
\subsection{Calculations for typical electroactive organic molecules}
The results of CASCI calculations are provided in Table~\ref{table:3mol}.
Here, the ionization energies and electron affinities decrease in the order of \textit{p}-benzoquinone $>$ thiohene $>$ pyrrole, corresponding to the trend exhibited by experimental data ~\cite{pyrrole_ie, thiophene_ie, p-bq_ie, pyrrole-thiophene_ea, p-bq_ea}.
These values were found to be lower in the case of the pyrrole and thiophene, which are more highly electron-donating, and higher for the more electron-accepting \textit{p}-benzoquinone.
This outcome is generally consistent with the known electronic properties of these molecules.
The calculated ionization energies were close to the experimental values although the electron affinities were underestimated.
A limited number of active orbitals was used in this study and the molecular orbitals were not optimized for each state, which may be responsible for the underestimation.
Another possible reason is the use of a basis function system that did not include a diffuse function, which is necessary to accurately describe the anionic state.
Because the focus of the present research was to examine the effect of \textit{w}, no attempts were made to improve the accuracy of these calculations.

\begin{figure}[H]
    \begin{center}
        \includegraphics[scale=0.85]{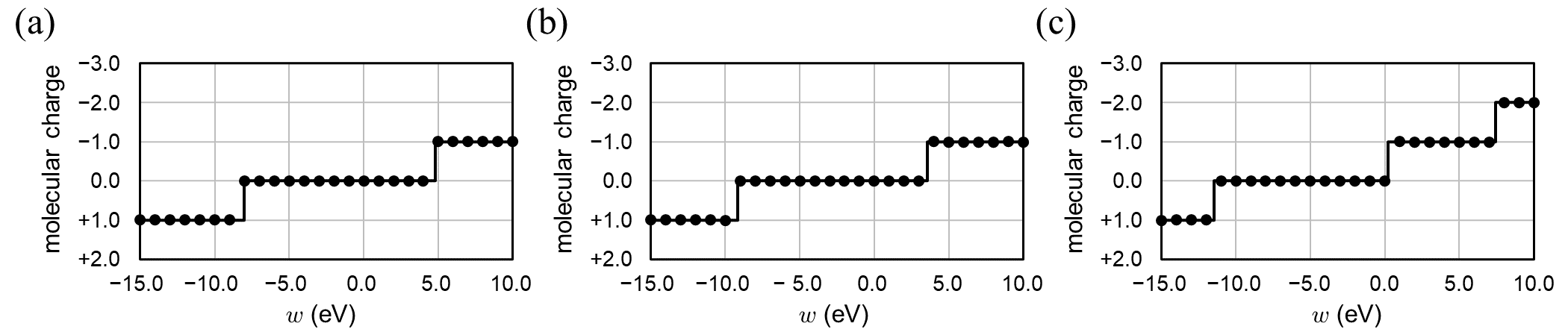}
        \caption{Variations in molecular charge for (a) pyrrole, (b) thiophene, and (c) \textit{p}-benzoquinone with changes in the weighting coefficient, \textit{w}.} 
        \label{3mol-chg}
    \end{center}
\end{figure}

\begin{figure}[H]
    \begin{center}
        \includegraphics[scale=0.85]{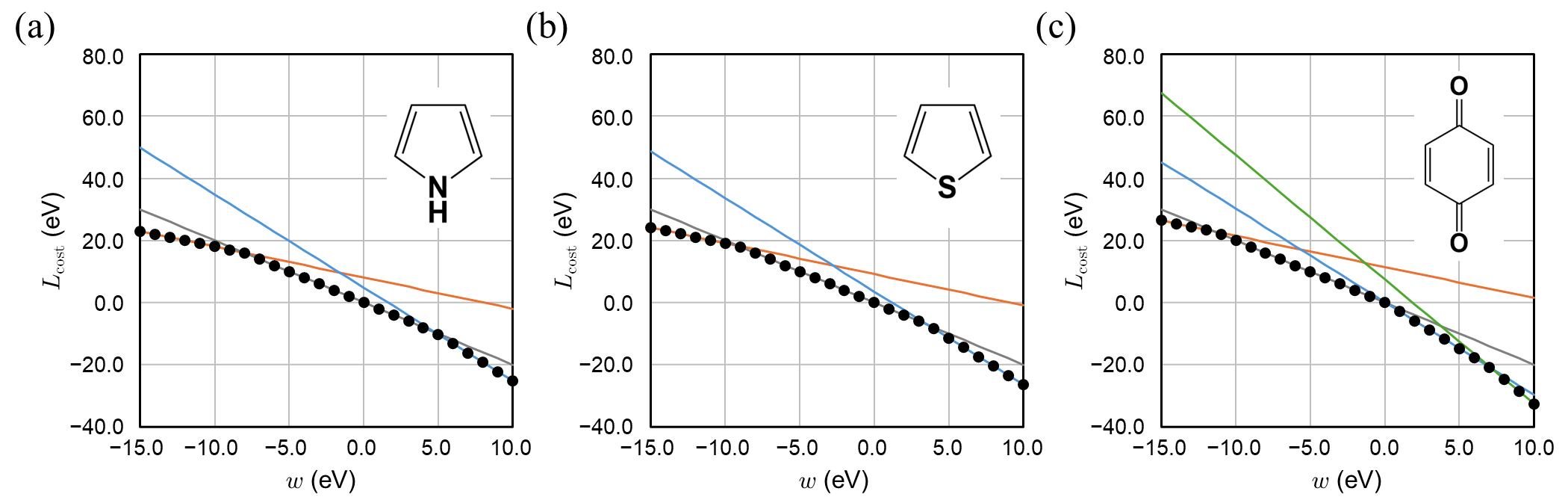}
        \caption{\textit{L}$_{\text{cost}}$ values obtained using the VQE method (filled circles). Values relative to those for neutral species with \textit{w} = 0.0 eV are plotted. \textit{L}$_{\text{cost}}$ values for neutral, cationic, anionic and dianionic species estimated from CASCI results are indicated by gray, orange, blue and green lines, respectively.}
        \label{3mol-Lcost}
    \end{center}
\end{figure}

Figure~\ref{3mol-chg} plots changes in molecular charge with respect to the calculated \textit{w} values generated using the VQE approach.
It is evident that, as \textit{w} was increased, the molecular charge changed from +1 to $\pm$ 0 then from $\pm$ 0 to $-$1.
The calculations for the more electron-accepting \textit{p}-benzoquinone indicated that a dianion having a charge of $-$2 appeared at $w \geq 8.0$ eV.
The \textit{w} values at which the molecular charge changes from \textit{i} to \textit{j} correspond to the $\Delta E_{i \rightarrow j}$ values summarized in Table~\ref{table:3mol}.
The \textit{L}$_{\text{cost}}$ values determined using the present VQE method are presented in Figure~\ref{3mol-Lcost} together with the \textit{L}$_{\text{cost}}$ for neutral states, cations, anions and dianions as estimated from the CASCI results.
For any given \textit{w}, the VQE calculations converged to the electronic state with the lowest \textit{L}$_{\text{cost}}$.
The ordering of the \textit{L}$_{\text{cost}}$ values for the states with charge \textit{i} and \textit{j} changes at $\Delta E_{i \rightarrow j}$ in Table~\ref{table:3mol}, in agreement with the variation in charge in Figure~\ref{3mol-chg}.
These results suggest that VQE calculations using a cost function as defined in eq \ref{lcost3} can involve variable electron numbers together with a constant electron chemical potential for a given \textit{w} in the case that an ansatz that does not conserve the electron number is employed.
This work also confirms that the molecular charge need not be specified a priori but can instead be considered as a physical quantity resulting from a calculation in which the solution converges to the state with the lowest \textit{L}$_{\text{cost}}$ for a given\textit{w}.
In addition, the \textit{w} at which the charge variation takes place corresponds to the vertical ionization energy or electron affinity.

\newpage
\begin{table}[H]
    \centering
    \caption{Electronic state energies for neutral, cationic, anionic and dianionic species, energy changes ($\Delta E$) with molecular charge variations and ionization energies (i.e.) and electron affinities (e.a.) for pyrrole, thiophene and \textit{p}-benzoquinone calculated using the CASCI method. The calculated second electron affinity value for \textit{p}-benzoquinone is also presented. Electronic state energies are given in hartree while all other values are in eV. Experimental values are in parentheses.~\cite{pyrrole_ie, thiophene_ie, p-bq_ie, pyrrole-thiophene_ea, p-bq_ea}}
    \footnotesize
    \begin{tabular}{lrrr} \hline
    \multicolumn{1}{c}{property}  &
    \multicolumn{1}{c}{pyrrole}   &
    \multicolumn{1}{c}{thiophene} &
    \multicolumn{1}{c}{\textit{p}-benzoquinone} \\ \hline \\
    \multicolumn{4}{l}{(electronic state energies in hartree)}\\ \\
    neutral  & $-$208.6594 & $-$551.0753 & $-$378.9453 \\
    cation   & $-$208.3642 & $-$550.7393 & $-$378.5245 \\
    anion    & $-$208.4838 & $-$550.9438 & $-$378.9380 \\
    dianion  &             &             & $-$378.6720 \\ \\
    \multicolumn{4}{l}{(energetic changes in eV)}\\ \\
    $\Delta E_{+1 \rightarrow \pm 0}$ & $-$8.03 & $-$9.14 & $-$11.45 \\
    $\Delta E_{\pm 0 \rightarrow -1}$ & 4.78 &    3.58 &     0.20    \\
    $\Delta E_{-1 \rightarrow -2}$    &         &         &     7.24 \\
    i.e. & 8.03 (8.21$^\text{a}$) & 9.14 (8.84$^\text{b}$) & 11.45 (9.99$^\text{c}$)      \\
    e.a. & $-$4.78 ($-$2.38$^\text{d}$) & $-$3.58 ($-$1.17$^\text{d}$) & $-$0.20 (1.89$^\text{e}$)    \\
    second e.a. & & & $-$7.24\\ \hline \\
    \multicolumn{4}{l}{$^\text{a}$ reference ~\cite{pyrrole_ie}}\\
    \multicolumn{4}{l}{$^\text{b}$ reference ~\cite{thiophene_ie}}\\
    \multicolumn{4}{l}{$^\text{c}$ reference ~\cite{p-bq_ie}}\\
    \multicolumn{4}{l}{$^\text{d}$ reference ~\cite{pyrrole-thiophene_ea}}\\
    \multicolumn{4}{l}{$^\text{e}$ reference ~\cite{p-bq_ea}}\\
    \end{tabular}
    \label{table:3mol}
\end{table}

\newpage

\newpage

\newpage

\subsection{Models for the adsorption of ammonia and water on Cu and CuO}

\subsubsection{Results of CASCI calculations}
\begin{figure}[H]
    \begin{center}
        \includegraphics[scale=0.85]{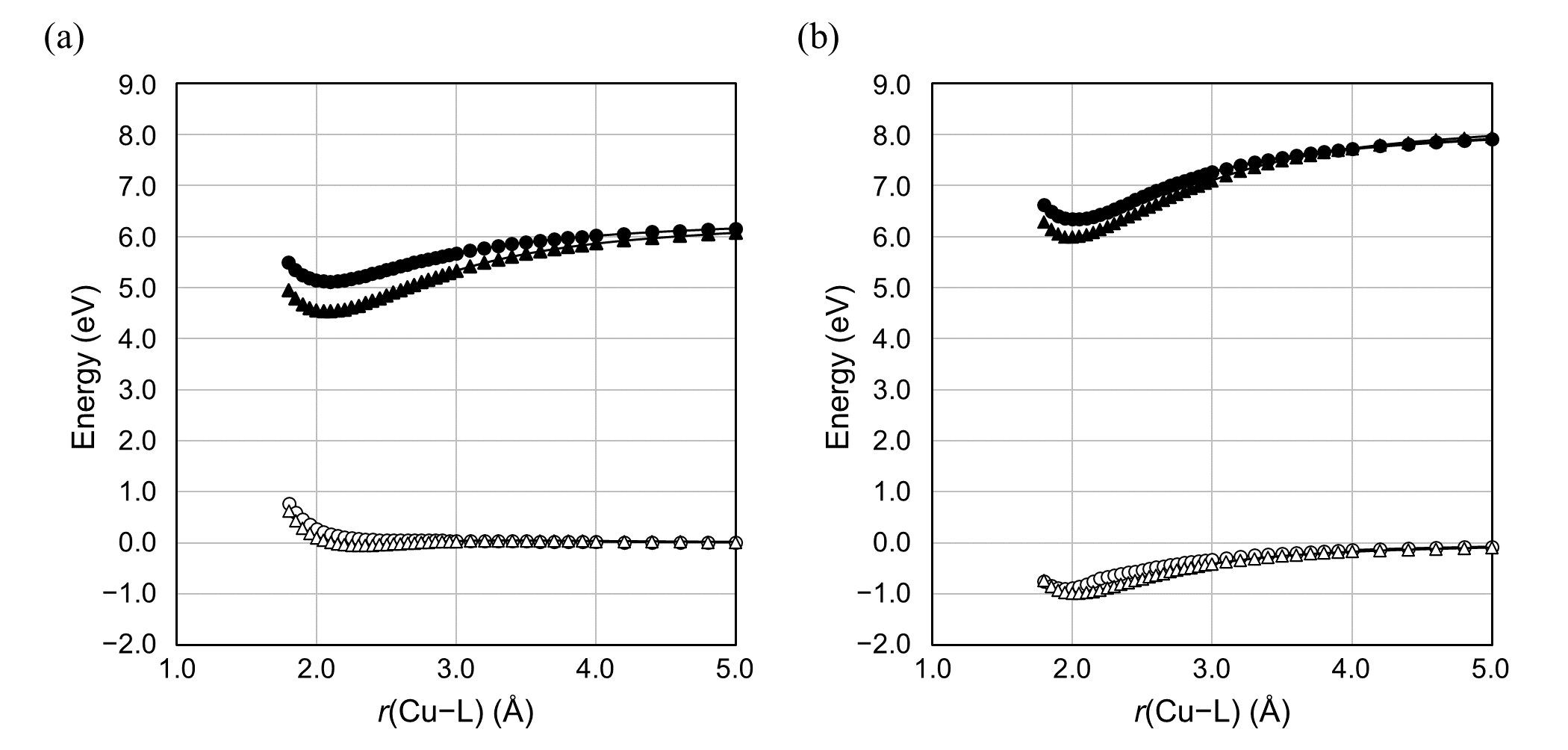}
        \caption{Electronic state energies of (a) Cu--X and (b) CuO--X models calculated using the CASCI method (X = H$_{2}$O, NH$_{3}$) for neutral species with water (open circles) and ammonia (open triangles) and cationic species with water (filled circles) and ammonia (filled triangles). For convenience, \textit{r}(Cu--O) and \textit{r}(Cu--N) are collectively denoted as \textit{r}(Cu--L). For each model, energies are plotted relative to neutral species with \textit{r}(Cu--L) = 20.0 \AA.}
        \label{cux-cuox_casci}
    \end{center}
\end{figure}

The results of CASCI calculations for Cu--X and CuO--X are shown in Figure~\ref{cux-cuox_casci}.
Both models suggest that a cationic species with a charge of +1 will interact more strongly than a neutral compound with a charge of $\pm$0 and that ammonia will interact more strongly than water.
The results for CuO--X produced a more deeply-bound curve than those for Cu--X, suggesting stronger interactions with X.
The energy difference between the neutral ([Cu--X]$^{0}$ and [CuO--X]$^{0}$) and cationic ([Cu--X]$^{+}$ and [CuO--X]$^{+}$) species was larger for CuO--X.
It should also be noted that the interaction in [Cu--X]$^{0}$ is primarily a result of dispersion forces, which will be stronger for the larger NH$_{3}$ molecule.
In addition, electrons occupying Cu 4s-derived orbitals will repel lone pairs in water and ammonia molecules.
Because the O atom of water has a greater $\delta^{-}$ charge than the N atom of ammonia as a result of its higher electronegativity, this repulsion is expected to be stronger in the case of [Cu--(H$_{2}$O)]$^{0}$.
Therefore, the interaction could be stronger for [Cu--(NH$_{3}$)]$^{0}$.
The present calculations generated a curve for [Cu--(NH$_{3}$)]$^{0}$ that was slightly attractive, whereas the curve for [Cu--(H$_{2}$O)]$^{0}$ was repulsive.
In contrast, in the case of the [Cu--X]$^{+}$, the Cu 4s electrons are removed to produce Cu$^{+}$, leading to a strong interaction with the lone pair electrons.
Since NH$_{3}$ has a lower ionization energy than H$_{2}$O, [Cu--(NH$_{3}$)]$^{+}$ will be more stable than [Cu--(H$_{2}$O)]$^{+}$.
The electrons on the Cu atom in the CuO--X are attracted to the bound O atom and so the Cu has a $\delta^{+}$ charge even in [CuO--X]$^{0}$.
Therefore, a stronger interaction should also occur in [CuO--X]$^{0}$ and is even more pronounced in [CuO--X]$^{+}$.
It is important to note that the CASCI and VQE calculations in this study dealt with a limited number of active orbitals and therefore did not account for the majority of the dynamic electronic correlations.
As a result, the accuracy of the quantitative calculations is expected to be low.
Nevertheless, as noted above, the shapes of the potential energy curves that were generated are those expected for the model compounds.

\newpage

\newpage
\subsubsection{Results of VQE calculations}
\begin{figure}[H]
    \begin{center}
        \includegraphics[scale=0.85]{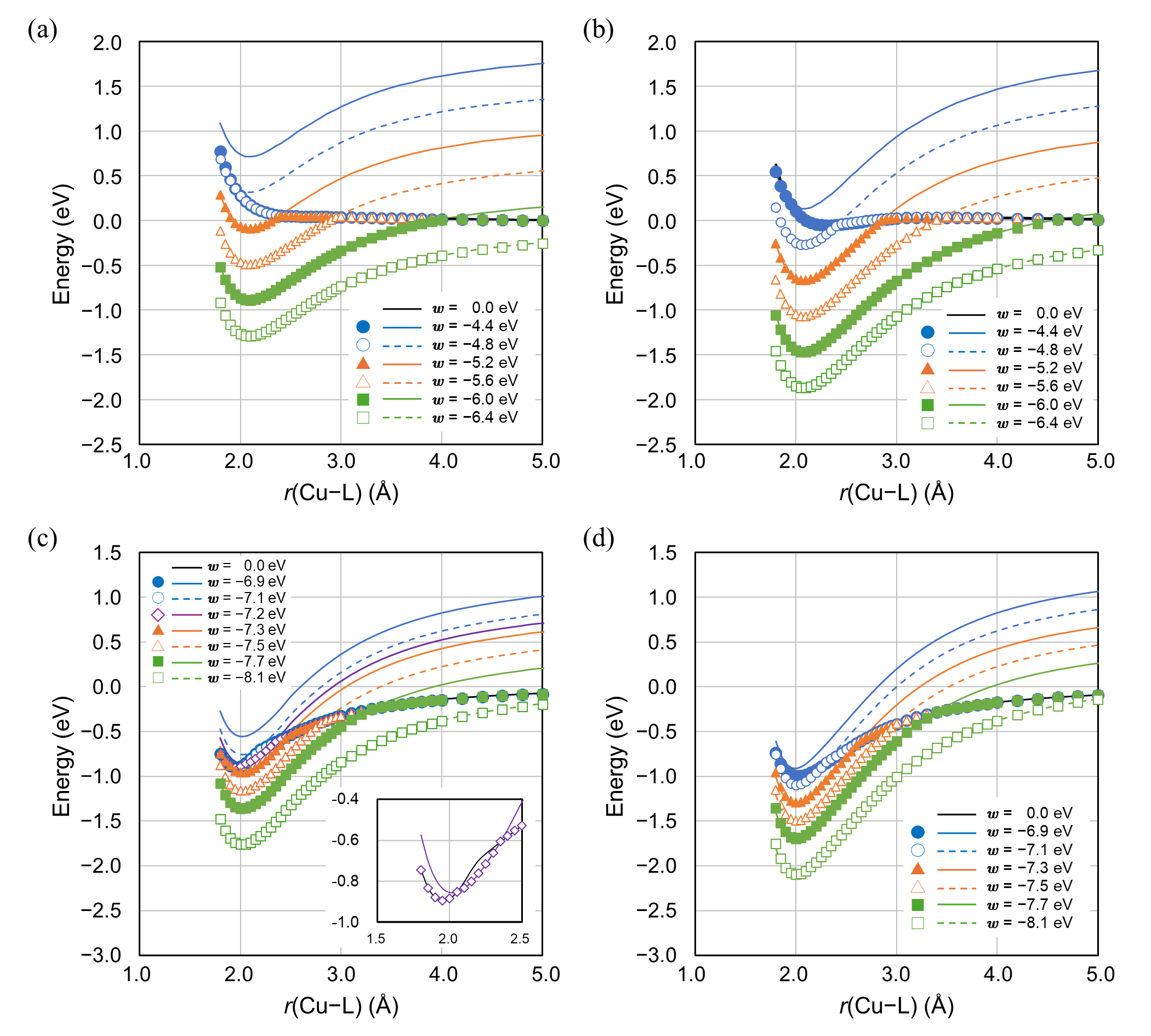}
        \caption{Electronic state energies for (a) Cu--(H$_{2}$O), (b) Cu--(NH$_{3}$), (c) CuO--(H$_{2}$O), and (d) CuO--(NH$_{3}$) models calculated using the CASCI (solid and dashed lines) and VQE (filled and open circles, triangles and squares) methods. For convenience, \textit{r}(Cu--O) and \textit{r}(Cu--N) are collectively denoted as \textit{r}(Cu--L). For each model, energies relative to neutral species with \textit{r}(Cu--L) = 20.0 \AA \: are plotted.}
        \label{vqe-wo-mc_cux-cuox_vqe}
    \end{center}
\end{figure}

Figure~\ref{vqe-wo-mc_cux-cuox_vqe} provides the results of the  \textit{L}$_{\text{cost}}$ calculations using the VQE method.
Here, the \textit{L}$_{\text{cost}}$ data are plotted relative to the values for the neutral species at \textit{r}(Cu--L) = 20.0 \AA.
The lines and dashed lines indicate the \textit{L}$_{\text{cost}}$ estimated based on CASCI calculations, on which the VQE calculations are plotted.
In most cases, there is a crossover of the \textit{L}$_{\text{cost}}$ curves for the neutral and cationic species, suggesting that the adsorption of X is accompanied by one-electron oxidation.
In the case of the Cu--(H$_{2}$O) data, the curve crossing occurs at $-$6.0 eV $\leq$ \textit{w} $\leq$ $-$4.8 eV.
The lower \textit{w} is,  the longer \textit{r}(Cu--L) the crossing occurs at.
At \textit{w} = $-$6.4 eV, there is no crossing and the cationic species are evidently stable over the entire range of \textit{r}(Cu--L) $\leq$ 5.0 \AA.
In addition, the relative energies of [Cu--(H$_{2}$O)]$^{+}$ decrease with decreases in \textit{w}, resulting in an increase in the adsorption energies.
The results for Cu--(NH$_{3}$) are similar to those for Cu--(H$_{2}$O) but also exhibit a crossing at \textit{w} = $-$4.4 eV, meaning that the cationic species should be more stable at \textit{r}(Cu--L) $\leq$ 1.95 \AA.
For $-$6.0 eV $\leq$ \textit{w} $\leq$ $-$4.8 eV, the crossing occurs at longer \textit{r}(Cu--L) than in the case of Cu--(H$_{2}$O).
As in the case of Cu--(H$_{2}$O), no crossing occurs at \textit{w} = $-$6.4 eV and the cationic species becomes stable throughout the entire range of \textit{r}(Cu--L) $\leq$ 5.0 \AA.
Very similar results to those for Cu--X were obtained for CuO--X although the \textit{w} value at which the curves crossed was lower than for Cu--X.
The crossover of the curves corresponds to the point at which the \textit{L}$_{\text{cost}}$ of the neutral species ($L^\text{neutral}_\text{cost}$) and that of the cationic species ($L^\text {cation}_\text{cost}$) are equal.
The latter will be lower in the region over which $-$\textit{w} is larger than the vertical ionization energy.
These values can be defined as

\begin{equation}
    \label{lcost-neutral}
    L^\text{neutral}_\text{cost} = E_\text{neutral}-w \cdot n_\text{neutral}
\end{equation}
and
\begin{equation}
    \label{lcost-cation}
    L^\text{cation}_\text{cost} = E_\text{cation}-w \cdot n_\text{cation},
\end{equation}

where $E_\text{neutral}$ and $n_\text{neutral}$ are the energy and electron number of the neutral species, respectively, and $E_\text{cation}$ and $n_\text{cation}$ are the energy and electron number of the cationic species, respectively.
In the case that $n_\text{neutral} - n_\text{cation}$ = 1, $L^\text{cation}_\text{cost} \leq L^\text{neutral}_\text{cost}$ will be true if

\begin{equation}
    \label{cation-neutral}
    E_\text{cation} - E_\text{neutral} \leq  -w,
\end{equation}

where the left-hand side is the vertical ionization energy.
Therefore, a lower value of \textit{w} will be associated with a wider range of \textit{r}(Cu--L) over which the cationic species are more stable.
In addition, since X = NH$_{3}$ has a lower ionization energy than X = H$_{2}$O, the cationic species will be more stable over a wider range for the same \textit{w}.
As well, in the case of CuO--X, which has a higher ionization energy, the \textit{w} at which the curves cross is lower than for Cu--X.

In all cases, the VQE calculations performed without charge specification converged to states having lower \textit{L}$_{\text{cost}}$, irrespective of the charge.
In the cases in which crossing of \textit{L}$_{\text{cost}}$ occurred, the calculations converged to neutral species in regions with longer \textit{r}(Cu--L), while to cationic species in regions with shorter \textit{r}(Cu--L).
The \textit{L}$_{\text{cost}}$ values at convergence were also in close agreement with those estimated using the CASCI process.
As a unique case, the calculation results for CuO--(NH$_{3}$) with \textit{w} = $-$7.2 eV are examined here.
As shown in the inset to Figure~\ref{vqe-wo-mc_cux-cuox_vqe}(c), in this scenario, the \textit{L}$_{\textit{cost}}$ crossing occurred at two structures: \textit{r}(Cu--N) = 2.05 \AA \ and \textit{r}(Cu--N) = 2.35 \AA. The cationic species were more stable in the range of 2.05 \AA $\leq$ \textit{r}(Cu--N) $\leq$ 2.35 \AA, whereas the neutral species were more stable in the other regions.
These findings suggest that the charge changes in the sequence of $\pm0 \rightarrow +1 \rightarrow \pm0$ upon adsorption.
The dual crossing points can possibly be attributed to the different shapes of the potential energy curves and to the \textit{r}(Cu--N) values associated with equilibrium for each state.
It was also observed that the electronic states with the lowest \textit{L}$_{\text{cost}}$ were once again successfully simulated using the VQE method.
In conclusion, the VQE method using eq \ref{lcost3} to define the cost function can ascertain the state having the lowest \textit{L}$_{\text{cost}}$ for a given \textit{w} while taking variations of the molecular charge into account.
Thus, quantum chemical calculations using quantum algorithms that can handle superpositions of electronic states with different numbers of electrons could be a useful means of analyzing chemical reactions with charge variations.

\newpage

\newpage
\section{Conclusion}

VQE is a quantum-classical hybrid algorithm capable of finding the electronic state for which a given cost function is minimized through optimization of variational parameters on a quantum circuit.
In this study, we apply an ansatz that does not conserve the number of electrons and demonstrated a VQE process in which the cost function is of the same form as the grand potential associated with the grand canonical ensemble.
In this framework, the weighting coefficients correspond to the chemical potentials of the electrons and the calculations converge to the electronic state that minimizes the cost function under those conditions.
The charge of the molecule is not specified in advance but rather is a physical quantity resulting from the calculations.
Calculations were performed for typical electron-donating molecules (pyrrole and thiophene) and an electron-accepting molecules (\textit{p}-benzoquinone) and yielded cationic, neutral or anionic species depending on the weighting coefficient.
Calculations involving the adsorption of water and ammonia on Cu and CuO suggested that one-electron oxidation can occur during the adsorption process depending on the weighting coefficients. The VQE method was also found to be capable of predicting the electronic state having the lowest cost function while considering variations in charge.
The results demonstrated the potential practical applications of this method.



\section{Author Information}

\textbf{Corresponding author}

\noindent
Soichi Shirai\\
Toyota Central Research and Development Laboratories, Inc.,\\
41-1 Yokomichi, Nagakute, Aichi 480-1192, Japan;\\
\url{https://orcid.org/0000-0001-6932-4845};\\
Email: \url{shirai@mosk.tytlabs.co.jp}

\noindent
\textbf{Authors}\\
\noindent
Takahiro Horiba\\
Toyota Central Research and Development Laboratories, Inc.,\\
41-1 Yokomichi, Nagakute, Aichi 480-1192, Japan;\\
\url{https://orcid.org/0000-0002-8610-047X}

\noindent
Hirotoshi Hirai\\
Toyota Central Research and Development Laboratories, Inc.,\\
41-1 Yokomichi, Nagakute, Aichi 480-1192, Japan;\\
\url{https://orcid.org/0000-0002-9618-3387}

\textbf{Notes}\\
The authors declare no competing financial interest.

\section{Acknowledgments}
The authors are grateful to Drs. Nobuko Ohba and Ryosuke Jinnouchi at Toyota Central R\&D Labs. Inc. for helpful discussions.

\bibliography{ref}

\newpage

\end{document}